# Product risk assessment: a Bayesian network approach

Joshua L. Hunte[a]*, Martin Neil[a,b] and Norman E. Fenton[a,b]

[a]School of Electronic Engineering and Computer Science, Queen Mary University of London, London, UK; [b]Agena Ltd, Cambridge, UK

*corresponding author: Joshua Hunte j.l.hunte@qmul.ac.uk

## ABSTRACT

Product risk assessment is the overall process of determining whether a product, which could be anything from a type of washing machine to a type of teddy bear, is judged safe for consumers to use. There are several methods used for product risk assessment, including RAPEX, which is the primary method used by regulators in the UK and EU. However, despite its widespread use, we identify several limitations of RAPEX including a limited approach to handling uncertainty and the inability to incorporate causal explanations for using and interpreting test data. In contrast, Bayesian Networks (BNs) are a rigorous, normative method for modelling uncertainty and causality which are already used for risk assessment in domains such as medicine and finance, as well as critical systems generally. This article proposes a BN model that provides an improved systematic method for product risk assessment that resolves the identified limitations with RAPEX. We use our proposed method to demonstrate risk assessments for a teddy bear and a new uncertified kettle for which there is no testing data and the number of product instances is unknown. We show that, while we can replicate the results of the RAPEX method, the BN approach is more powerful and flexible.

# 1. Introduction

It is essential that the products we use in our homes are acceptably safe. To ensure our safety, national regulators perform product risk assessments to limit consumer harm (RPA 2006; PROSAFE 2013; European Commission 2015, 2018). There are several methods used for product risk assessment such as Nomograph (RPA 2006), Matrix (RPA 2006) and RAPEX (RPA 2006; European Commission 2015, 2018), but these methods vary in how they quantify risks (European Commission 2018). In this paper, we identify a number of limitations of RAPEX in particular and explain the need for a systematic method for product risk assessment that: takes full account of uncertainty; uses causal knowledge of both the testing and operational environment and the process by which data are generated; considers the user population at risk and the product risk tolerability. We propose that Bayesian networks (BNs) can provide such a systematic method as they are a rigorous, normative method for modelling uncertainty and causality. In fact, BNs are already used for risk assessment in a range of critical applications domains. For example, medical (Fenton and Neil 2010; Kaewprag et al. 2017; M. Li et al. 2019), transport (Marsh and Bearfield 2004; Baksh et al. 2018; Russo et al. 2017), nuclear (C. J. Lee and Lee 2006), system reliability and dependability (Weber et al. 2012), construction (Leu and Chang 2013; Zhang et al. 2014), technology (S. Li et al. 2016; Wang, Neil, and Fenton 2020), project management (Chin et al. 2009; E. Lee, Park, and Shin 2009) and finance (Neil, Häger, and Andersen 2009; Masmoudi, Abid, and Masmoudi 2019).

There has also been some previous work on using BNs for product risk assessments. For instance, Suh (2017) developed a product risk assessment system using a BN to assess product risk based on injury information from the Korea Consumer Agency. They evaluated 33 children's products and compared the results with RAPEX. Berchialla et al. (2012) used a BN to estimate the risk of ingestion, inhalation and insertion of consumer products in children aged 0-14. They also compared the BN approach to other quantitative risk assessment methods such as neural networks, classification trees, and logistic models (Berchialla et al. 2016). Their results indicate that BNs are the best method for assessing the risk of ingestion, inhalation and insertion of consumer products in children due to their ease of interpretability and prediction accuracy (Berchialla et al. 2016). However, these previous works have not explicitly described or proposed a generic BN for product risk assessment. For instance, the Berchialla et al. (2012) BN cannot be used for product risk assessment since its structure and parameters are not applicable. Though Suh (2017) used a BN in their product risk assessment system, it is unclear if this BN is suitable since they have not provided a causal diagram (a diagram that describes the causal relationship between the variables in the BN) and any description of the BN is written in Korean. We present in this article a generic BN that significantly extends the previous work on BNs for product risk assessment. The proposed BN incorporates hazard and injury data, manufacturer process information, product usage data and consumer utility and risk perception to estimate product risk.

The rest of this article is organised as follows: Section 2 provides an overview of product risk assessment, identifying the limitations of RAPEX. Section 3 describes our proposed BN model for product risk assessment. Section 4 describes two case studies using our proposed BN model. Finally, our conclusion and recommendation for further work are presented in Section 5.

**2. Product risk assessment overview**

A product is any item offered in a market to meet consumer needs. A product risk assessment is the overall process of determining whether a product, which could be anything from a type of washing machine to a type of teddy bear, is safe for consumers to use. Specifically, it is the process by which the level of risk associated with a particular (product) hazard is identified and categorised. The risk assessment process includes risk analysis and risk evaluation (European Commission 2015; ISO/IEC 2014):

(1) *Risk Analysis:* This phase involves hazard identification and risk estimation (ISO/IEC 2014).

   a. *Hazard identification:* The process of finding, recognising, and describing product hazards (European Commission 2015, 2018; ISO/IEC 2014). Hazards are potential sources of harm or injury and are intrinsic to the product (European Commission 2015, 2018; ISO/IEC 2014).

   b. *Risk Estimation:* The process of determining the risk level of the product (European Commission 2015, 2018; ISO/IEC 2014). Risk is the combination of the likelihood of a hazard causing injury to a consumer and the severity of that injury (European Commission 2015; ISO/IEC 2014). The risk level is the degree of the product risk on a scale from 'low' to 'serious' (European Commission 2015, 2018).

(2) *Risk Evaluation:* The process by which the outcome of the risk analysis is combined with policy considerations to characterise the risk and inform decisions on risk management (European Commission 2015). It includes determining if the risk is acceptable (European Commission 2015; ISO/IEC 2014).

As RAPEX (RPA 2006; European Commission 2015, 2018) is the most widely used method for product risk assessment, this article will review the RAPEX method (RPA 2006; European Commission 2015, 2018) and its limitations.

*2.1 RAPEX overview*

RAPEX was developed for the rapid exchange of information between the Member States of the European Union on the measures and actions concerning products that pose a serious risk to the safety and health of consumers (European Commission 2015, 2018). An essential component of RAPEX is product risk assessment which determines product risk and informs risk management response (European Commission 2015, 2018). The following steps or guidelines describe the method used by RAPEX for product risk assessment:

(1) *Describe the product*: Product details such as name, brand and model are documented during this stage (European Commission 2015, 2018).

(2) *Describe product hazards*: Hazards are identified by tests and standards or by the manufacturers' product labelling and instructions. A product may have one or more hazards such as thermal, electrical, and mechanical (European Commission 2015, 2018).

(3) *Identify consumers at risk*: Consumer types include intended users, non-intended users and vulnerable users (European Commission 2015, 2018).

(4) *Describe the injury scenario*: Injury scenarios describing how the product hazard may harm the consumer are developed (European Commission 2015, 2018). For example, an axe breaks and the ejected part strikes the user's head.

(5) *Determine the probability of injury*: Probabilities are assigned to each step of the injury scenario to determine the probability of injury (European Commission 2015, 2018). For example, to determine the probability of injury while using an axe, we combine the following probabilities:

a. Probability of axe breaking = 1/100

b. Probability of broken part hitting body = 1/10

c. Probability of broken part hitting head = 1/10

Total probability of injury = 0.01 * 0.1* 0.1 = 0.0001

Probabilities used in this step are assumed to be independent and are obtained from what are assumed to be reliable sources such as the European Injury Database and hospital injury databases (European Commission 2015, 2018).

(6) *Determine the severity of the injury*: The severity of the injury is determined by the type of medical intervention required for the injury scenario (European Commission 2015, 2018). The injury severity level and associated medical intervention are shown in Table 1.

| Injury Severity Level | Medical Intervention |
|---|---|
| 1 | First Aid |
| 2 | Visit Accident and Emergency Department (A&E) |
| 3 | Hospitalisation |
| 4 | Fatal or loss of a limb(s) |

Table 1. Injury severity level and associated medical intervention

For example, we assign a severity level of 2 for the injury scenario "an axe breaks and the ejected part strikes the user's head", since it may require a visit to A&E.

(7) *Determine the risk level*: This is determined by using a risk level matrix that combines the severity of the injury and the probability of the injury occurring described in the injury scenario (European Commission 2015, 2018). The estimated risk level of the product will contain some level of uncertainty since the probability of injury and severity of injury are estimated parameters. RAPEX handles uncertainty in the estimated risk level using a sensitivity analysis which determines how variations in the estimated parameters (i.e. probability of injury and the severity of injury) affect the overall risk result (European Commission 2015, 2018). It entails repeating the risk assessment process using different probabilities for the steps in the injury scenario and different injury severity levels. If the sensitivity analysis shows that there is no significant change in the risk level, then there is increased confidence in the initial estimated risk level. In contrast, a significant change will reduce confidence and require a review of the estimated parameters (European Commission 2015, 2018). For instance, if the initial risk assessment for an axe determines that the risk level is 'low' and the sensitivity analysis also shows that there is no significant change in the risk level, then the risk level of the axe is confidently considered as 'low'.

Despite the widespread use of the RAPEX method, it has the following limitations:

(1) *Limited approach to handling uncertainty*: In RAPEX, probabilities are assigned using point values instead of distributions. RAPEX handles second-order uncertainty (i.e. the uncertainty in the estimation of the parameters of interest (Briggs et al. 2012)) using a sensitivity analysis which entails repeating the risk assessment process using different probabilities for the steps in the injury scenario and different injury severity levels. This method of handling uncertainty is not practical for probabilities that are not directly observable, nor uncertainty about the data themselves due to the large number of possible probabilities that can be assigned. Nor can it handle uncertainty associated with novel products (i.e. those for which little or no relevant historical data exist) or products for which limited testing data are available.

(2) *Does not incorporate causal explanations for using and interpreting the data*: RAPEX provides no systematic or rigorous method for taking account of causal knowledge and explanations of the statistical data it uses, which may lead to inaccurate results. The most general example is that lack of incident data for a product may be due to lack of reporting on the product rather than lack of incidents while, at the other extreme, multiple incidents associated with a product may be the result of testing the product beyond its intended scope.

(3) *Does not differentiate between different types of users – i.e. their usage profile and risk tolerability*: In the RAPEX method, product risk is based on the likelihood of a product causing injury to a 'generic' user and the severity of that injury without any consideration of the context of use (RPA 2006; European Commission 2015, 2018). Hence, a product formally classified as 'high risk' may actually be 'low risk' or 'tolerable' for different classes of users taking account of the way they use the product, the benefits they receive from it and risk controls and mitigants. Risk controls and mitigants vary for different types of users due to their knowledge of the hazard and the environment they use the product. For instance, users that are aware of a fire hazard from a device are likely to have a smoke alarm installed nearby thus reducing the likelihood of injury, e.g. burn, even if the hazard occurs.

(4) *Does not consider the user exposure to the risk*: RAPEX does not include the usage frequency when determining the probability of a product causing injury to a user. Usage frequency is essential to determining the probability of injury since injury can only occur during product use. For instance, a consumer that uses a product often will have a higher probability of being injured due to repeated exposure to the hazard when compared to a consumer that rarely uses the product.

(5) *Does not include information on risk tolerability*: Risk tolerability is the trade-off between risk and utility. For instance, a 'high risk' product may be considered 'tolerable' for some users since they value the utility of the product sufficiently high and are willing to tolerate the 'high risk' as a trade-off for the utility. Hence, risk tolerability is an essential component of product risk assessment since it informs risk management response to a non-compliant product.

(6) *Does not consider increased risk of hazards over the lifetime of a product*: Due to wear and tear the 'hazard rate' of a product will generally increase over time, with different classes of products having very different increasing hazard rates. An estimated hazard rate of a product – based only on testing instances of the product when new – will underestimate the true hazard rate of the product in operation.

## 3. The Bayesian network (BN) model

Bayesian networks (BNs) are a type of probabilistic graphical model that explicitly describe dependencies between a set of variables using a directed acyclic graph (DAG) and a set of node probability tables (NPTs) (Fenton and Neil 2018; Pearl 2009; Spohn 2008). The directed acyclic graph consists of nodes and directed arcs; nodes represent variables and arcs are used to link nodes (Fenton and Neil 2018; Spohn 2008). The arcs assume that there is a causal influence or statistical relation between nodes (although in our work we generally assume the relationship is causal, and emphasise this by referring to them as causal BNs). For instance, given two nodes X and Y, an arc from X to Y assumes that Y directly influences X; as a result, X is called the parent of Y and we also say Y is dependent on X (although note that this means statistical dependence – it does not mean it is the only thing that can 'cause' Y). Each node in a DAG has a node probability table (NPT) which describes the probability distribution of the node conditional on its parents. Any node without a parent is called a root node and the NPT for that given node is its probability distribution (Fenton and Neil 2018).

Given the limitations of RAPEX discussed in Section 2.1, we propose the BN model for product risk assessment shown in Figure 1. We used the knowledge-based and idioms-based approaches for model development. The knowledge-based approach entails using domain expert knowledge to build the BN structure and to define model parameters where data are limited or missing (Constantinou and Fenton 2018). The idioms-based approach entails using idioms to build the BN structure. Idioms are fragments of a BN that represent generic types of uncertain reasoning (Neil, Fenton, and Nielsen 2000).

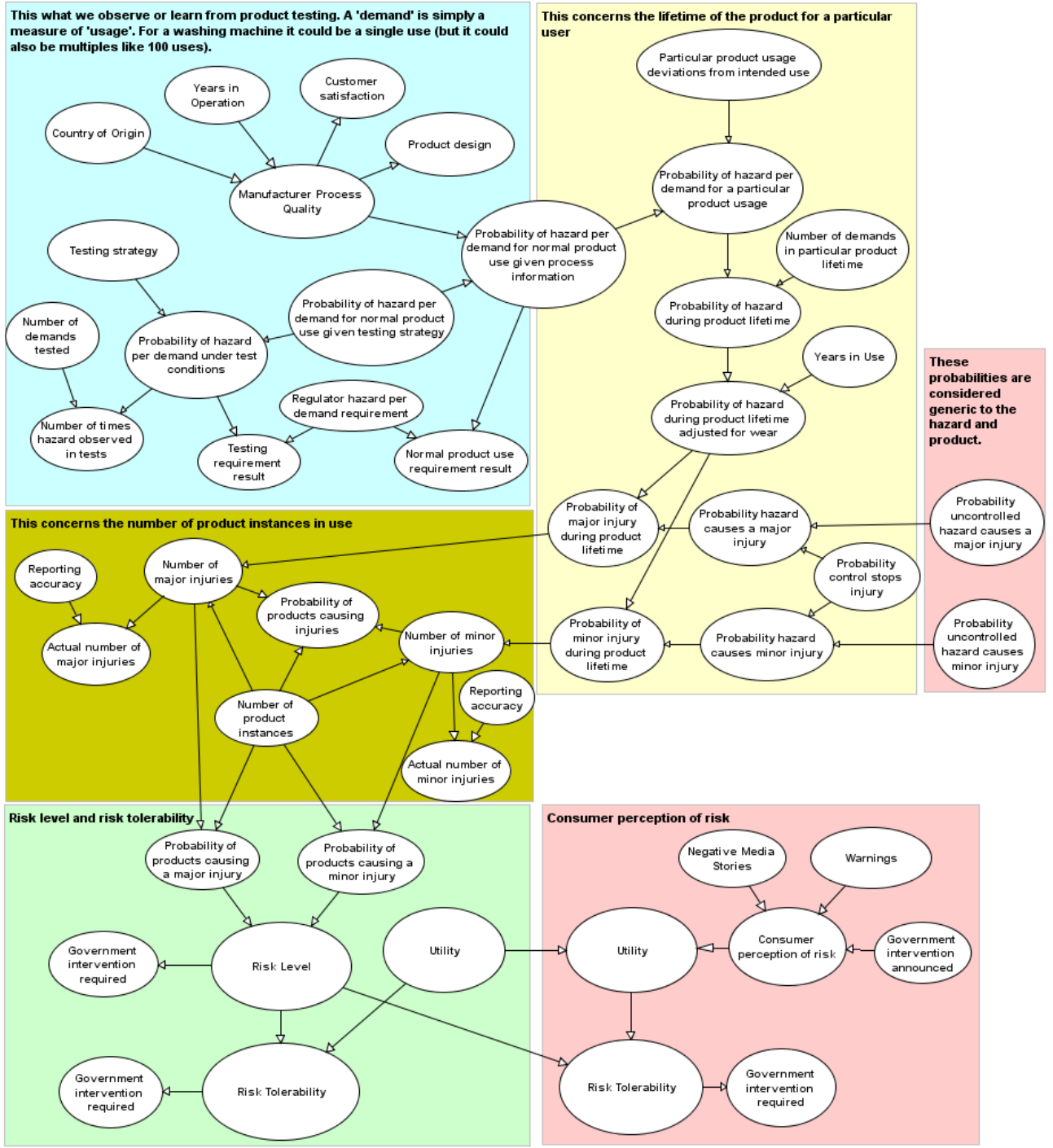

Figure 1. Bayesian network model for Product risk assessment

Our proposed BN model resolves the following issues with the RAPEX method:

(1) *Properly handles uncertainty about probabilities assigned during risk assessment*: Our BN model handles second-order uncertainty by incorporating distributions, rather than point values for probabilities that are not directly observable.

(2) *Incorporates causal explanations for using and interpreting the data*: Our BN model explicitly describes the risk assessment process and the causal relationship between the data used.

(3) *Considers the usage behaviour for different types of users and the number of product instances when determining the product risk*: The BN model can take full account of the distributions of different types of users when estimating product risk by simply assigning priors to the 'particular product usage' node that capture the population distribution. For instance, if for a particular product we estimate that only 30% of the population will 'use it as intended' then we set the prior probability of that node state to 30%. In addition, the BN model explicitly includes 'controls' that can prevent a hazard from causing an injury. For example, in households where there is a smoke alarm and fire extinguisher, the probability that a fire from a washing machine leads to injury is greatly reduced. In households where young children are under close supervision, there is a much lower probability that a hazard from a toy (such as an eye pulled off a teddy bear) would lead to injury compared to households where children are left unsupervised. Lastly, the BN model can provide individualised risk assessments. For instance, for a particular user, the model can estimate the probability that this user will suffer an injury during the product lifetime.

(4) *Considers the user exposure to risk:* Our BN model uses the usage frequency of the product to determine the probability of injury for a particular user or class of user.

(5) *Includes information on risk tolerability*: Our BN model combines utility and the product risk to determine risk tolerability for a particular user or class of user.

(6) *Considers increased risk of hazards over the lifetime of a product:* Our BN model considers the effect of wear and tear on the 'hazard rate' of the product.

(7) *Estimates the risk for novel products (i.e. those for which little or no relevant historical data exist) or products for which limited testing data are available:* In situations where it will neither be feasible nor possible to get any extensive data from testing, our BN model can use expert judgement and/or estimates from previous similar products together with process information about the manufacturer and their reputation to estimate product risk.

Our BN model also improves product risk assessment by modelling:

(1) *The effect of a Government intervention, negative media stories and warnings on the consumer perception of the risk*: For instance, as shown in Figure 2, if there are little or no negative media stories, no warnings, and no Government intervention announced for a particular product, the BN model predicts no change in consumer perception of the risk. However, as shown in Figure 3, if there are negative media stories, warnings and a Government intervention is announced, the BN model predicts the consumer perception of the risk will most likely change.

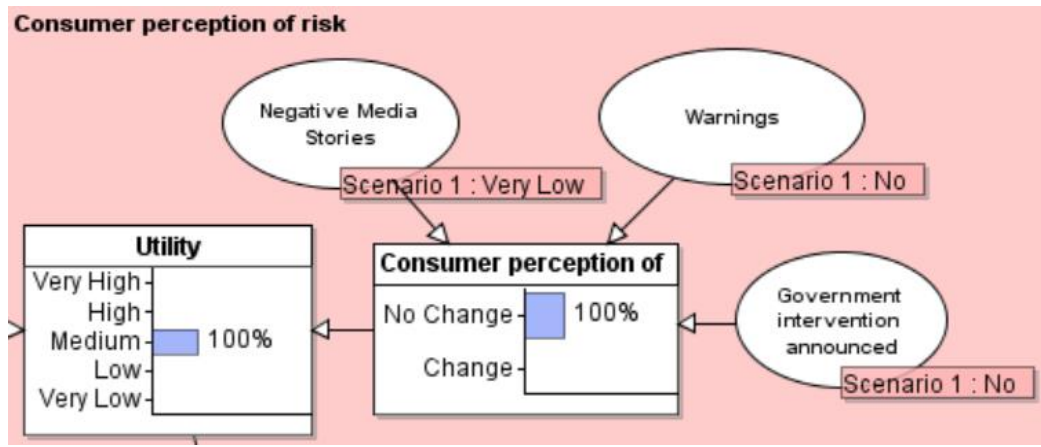

Figure 2. BN fragment showing no change in consumer perception of risk

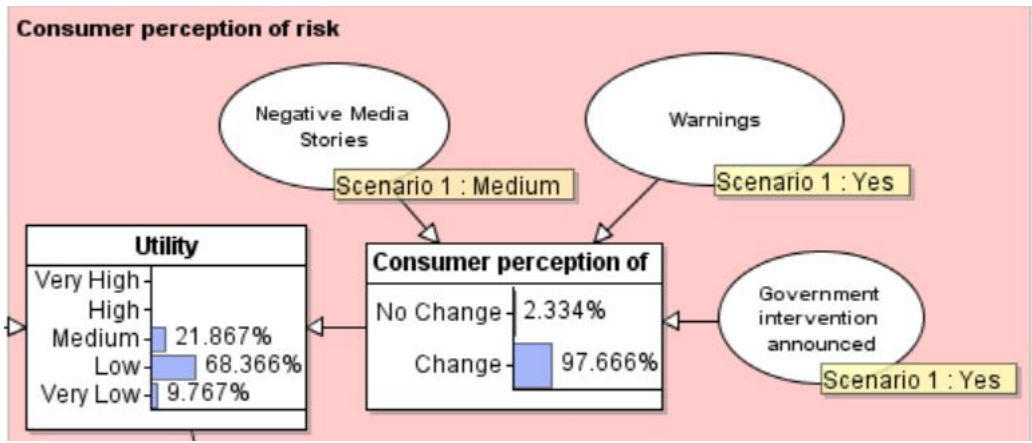

Figure 3. BN fragment showing the change in consumer perception of risk

(2) *The mean number of product instances causing major and minor injuries respectively*: For instance, as shown in Figure 4, the BN model can estimate the mean number of product instances causing major and minor injuries for a particular product based on the total number of product instances and the probability distribution of major and minor injuries respectively. In this example, for a particular product with 519,000 instances and mean probabilities of causing major and minor injuries as 0.018 and 0.036 respectively for a particular user or class of user, the BN model estimate that the mean number of product instances causing major and minor injuries are 9335 and 18668 respectively.

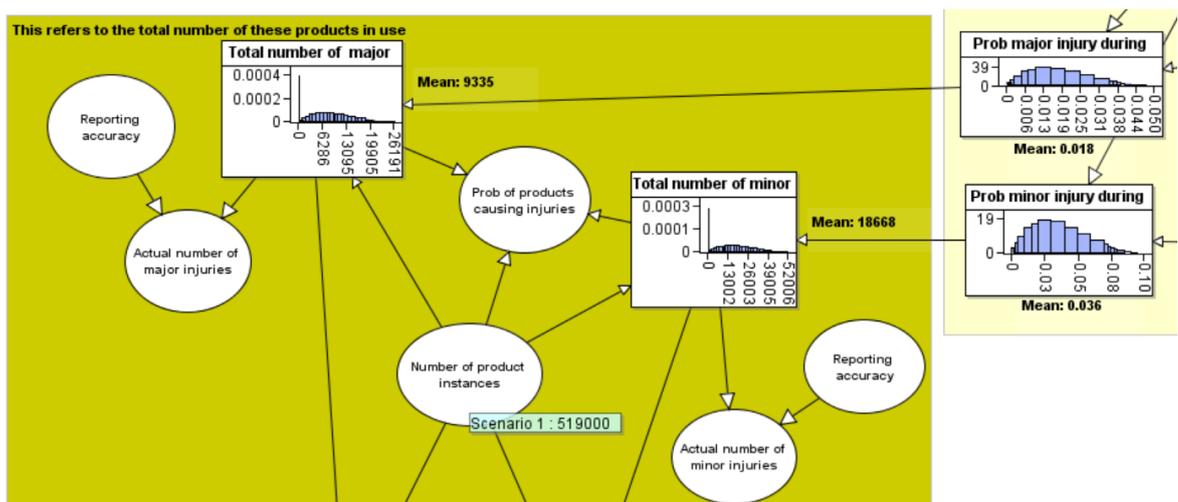

Figure 4. BN fragment showing the mean number of product instances causing major and minor injuries

(3) *The probability of hazard per demand using testing data and process information about the manufacturer and its reputation*: For instance, as shown in Figure 5, the BN model can estimate the probability distribution of the hazard per demand for normal product use based on the probability distribution of hazard per demand observed during testing. In this example, for a particular product with 2000 demands, 1 observed hazard and a testing strategy 'typical of normal use', the BN model estimates that the probability distribution of hazard per demand for normal product use (mean 0.001) is the same as the probability distribution of hazard per demand observed during testing (mean 0.001). If product testing was 'poor' as shown in Figure 6, the probability distribution of hazard per demand for normal product use (mean 0.02) varies significantly from the distribution observed during testing (mean 0.001). This variation is due to Bayes' inference which revises the probability distribution of the hazard per demand for normal product use based on the testing strategy.

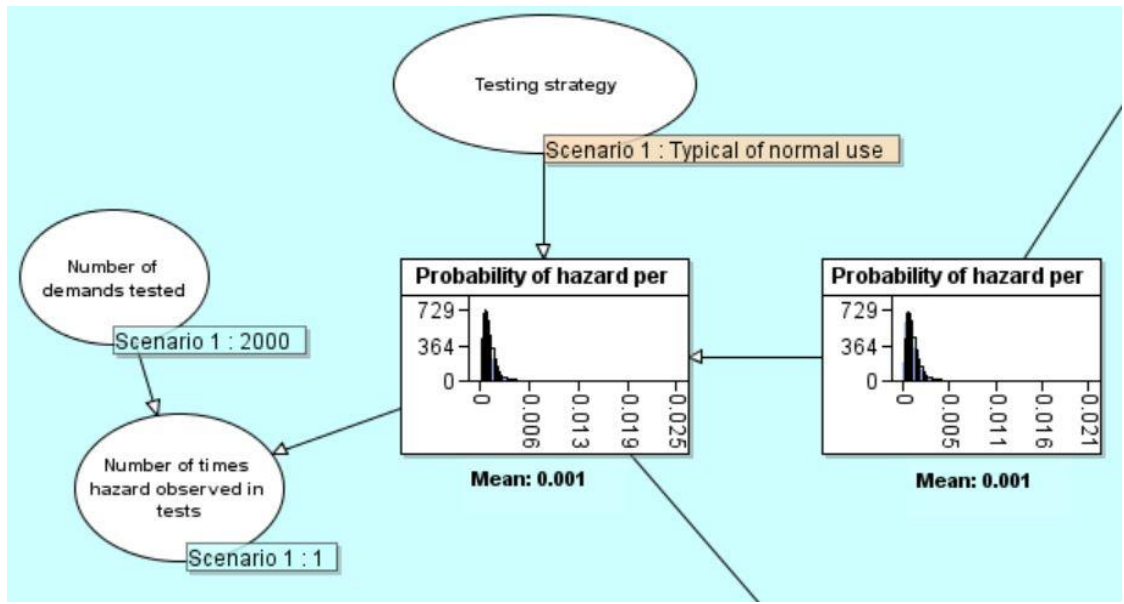

Figure 5. BN fragment showing the probability distribution of hazard per demand for a particular product with a testing strategy typical of normal use

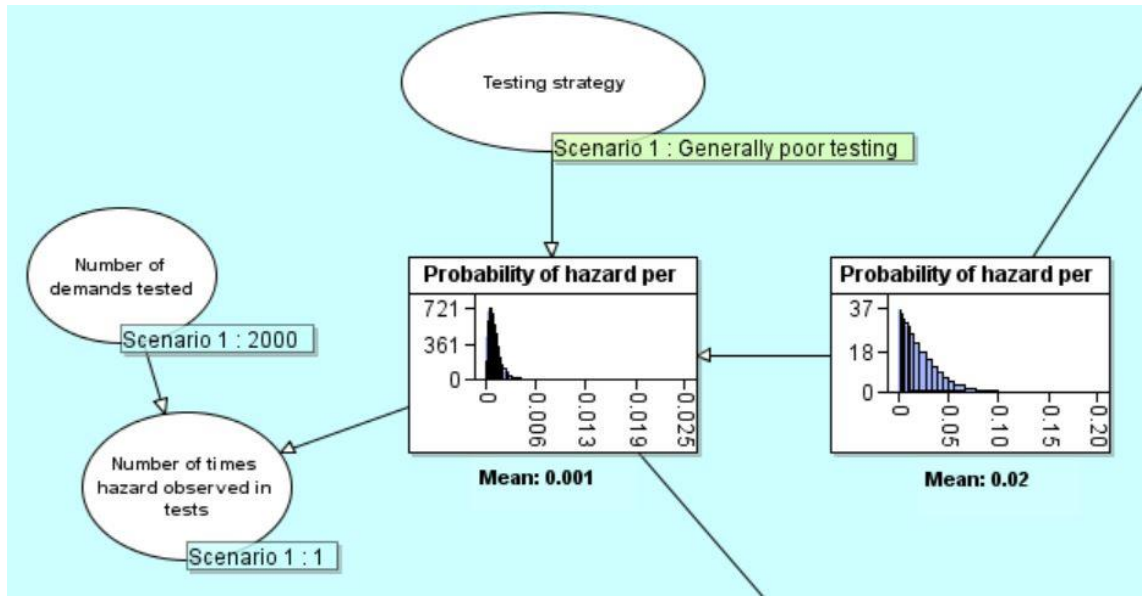

Figure 6. BN fragment showing the probability distribution of hazard per demand for a particular product with a poor testing strategy

## 4. Case study results

In this section, we will use our proposed BN model to determine the risk associated with a Teddy Bear (Section 4.1) and a new uncertified kettle (Section 4.2).

### 4.1 Teddy Bear risk assessment using the BN model

#### 4.1.1 Background

Every parent is concerned about toy safety. In the UK and the EU, the RAPEX method is used to assess the risk associated with different toys to prevent harm to children. We will use our BN model to assess the risk associated with a teddy bear using two scenarios and compare our results to the RAPEX method.

#### 4.1.2 Scenarios

We assume that the hazard is a small part, e.g. a teddy bear eye, which is swallowed by a child resulting in an injury.

(1) *Scenario 1:* We assume that the teddy bear is used by a child aged 0-36 months as intended for 1 year with a high number of demands (i.e. 4000) and no parental intervention.

(2) *Scenario 2:* We assume that the teddy bear is used by a child aged 0-36 months as intended for 1 year with a low number of demands (i.e. 200) and an 80% chance of parental intervention.

We also assume that the probability of the uncontrolled hazard causing a major injury 0.1 and a minor injury 0.2. We assume that the product was tested 'typical of normal use' and for 5000 demands we observe 1 hazard. We assume that there are 20,000 product instances and the utility of the teddy bear is 'medium'. Finally, we assume that the manufacturer has been in operation for 5-10 years and is from a country with a good safety record for toys. The manufacturer also has a 'high' customer satisfaction rating, and there are no changes in product design (i.e. product appearance is the same as previous similar products).

*4.1.3 Results*

(1) *Scenario 1*: Our BN model shown in Figure 7 learns that the risk level for the teddy bear is 'very high' with little uncertainty for this particular user or class of user. Our BN model also calculates that the mean probability of a major injury is 0.07 and a minor injury is 0.14. It also shows that the mean number of major and minor injuries for 20,000 product instances is 1387 and 2773 respectively. Finally, the BN model shows that the risk tolerability distribution for the teddy bear is in the range of 'low' to 'very low' given a 'medium' utility and that a Government intervention such as recall is required. One of the limitations of the RAPEX method is that it does not consider the number of demands for a particular product when determining risk. So, although we are unable to make a direct comparison to our BN model, we can compare the product risk result of our BN model to the RAPEX method by using the mean probability of a major injury learnt by our BN model as the probability of injury for the RAPEX system. We set the injury severity level to '3' as this corresponds to a major injury such as internal airway obstruction. The RAPEX method assesses the risk level of the teddy bear as 'serious' as shown in Figure 9. This result is the same as our model even though our model also uses the probability of a minor injury to compute the product risk.

(2) *Scenario 2*: Our BN model shown in Figure 8 learns that the risk level for the teddy bear is centered at 'low' to 'very low' with some uncertainty for this particular user or class of user. Our BN model also calculates that the mean probability of a major injury is 0.001 and a minor injury is 0.003. It also shows that the mean number of major and minor injuries for 20,000 product instances is 35 and 68 respectively. Finally, the BN model shows that the risk tolerability distribution for the teddy bear is centered at 'high' given a moderate utility and recommends no Government intervention such as a recall with some uncertainty. One of the limitations of the RAPEX method is that it does not consider the number of demands for a particular product when determining risk. So, although we are unable to make a direct comparison to our BN model, we can compare the product risk result of our BN model to the RAPEX method by using the mean probability of a major injury learnt by our BN model as the probability of injury for the RAPEX system. We set the injury severity level to '3' as this corresponds to a major injury such as internal airway obstruction. The RAPEX method assesses the risk level of the teddy bear as 'serious' as shown in Figure 10. This result is not the same as our model for the given probability of a major injury since our model also uses the probability of a minor injury to compute the product risk.

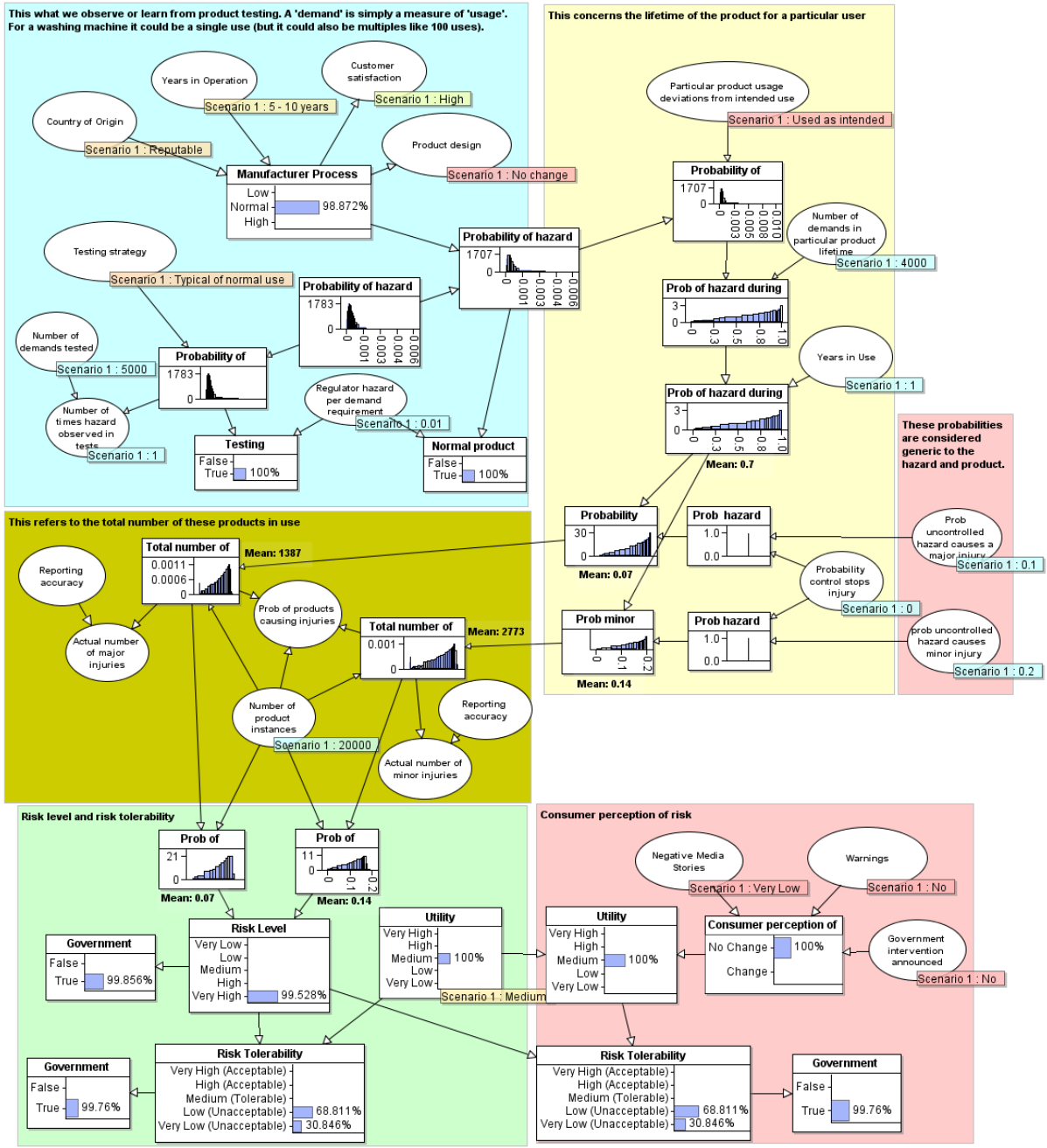

Figure 7. BN Model for Teddy bear Scenario 1

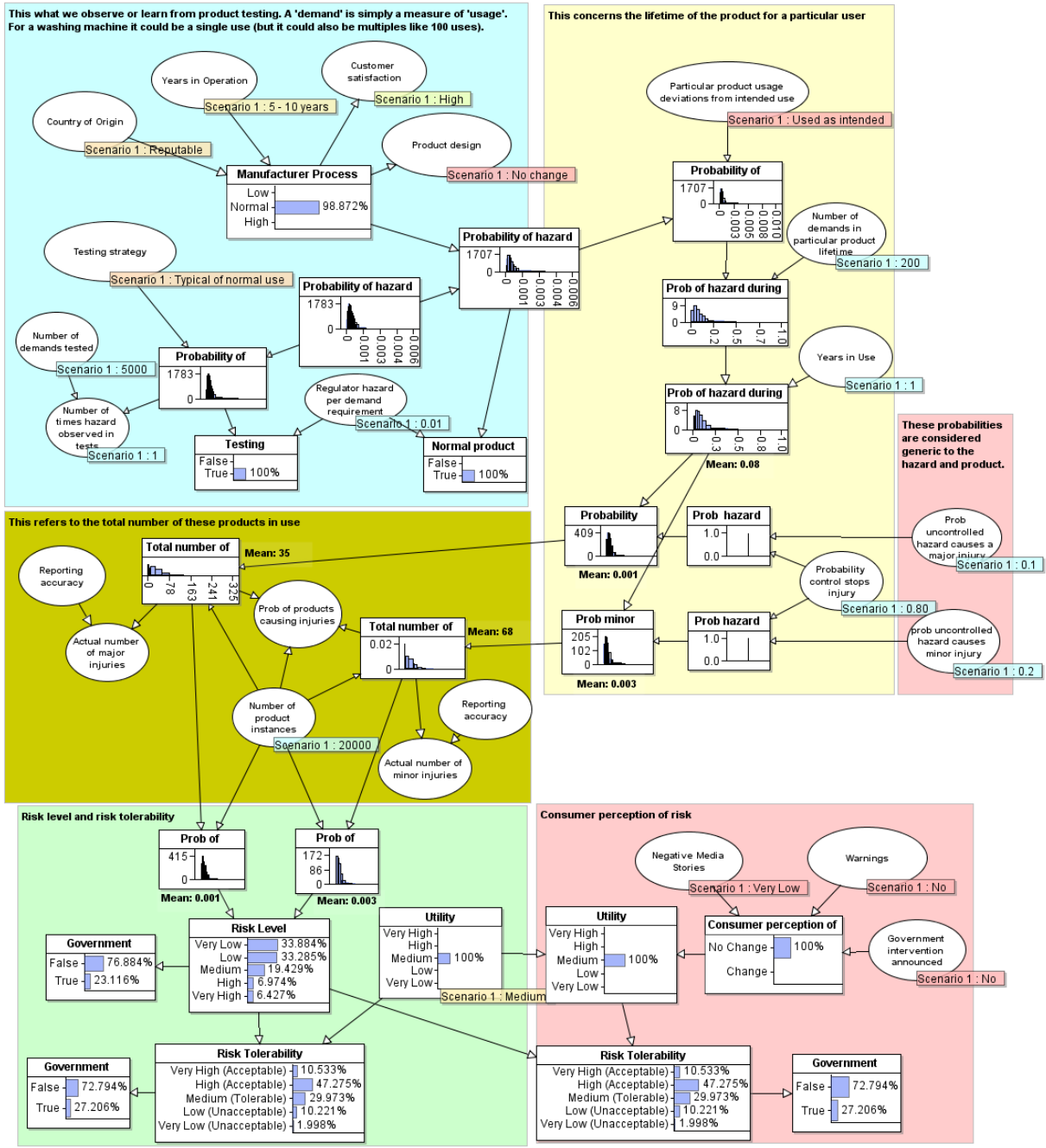

Figure 8. BN Model for Teddy bear Scenario 2

Figure 9. RAPEX results for Teddy bear Scenario 1 (European Union 2019)

Figure 10. RAPEX results for Teddy bear Scenario 2 (European Union 2019)

*4.2 A new uncertified Kettle risk assessment using the BN model*

*4.2.1 Background*

Every year there are new uncertified products available on the market that pose a serious risk to the health and safety of consumers. However, regulators are unable to assess the risk for these products using RAPEX, since there is no testing data, and the number of product instances is unknown. We will use our BN model to assess the risk for two different scenarios, both involving a new uncertified kettle on the market for which there are no testing data, and the number of product instances is unknown. The hazard is an ignition source which causes a fire resulting in a burn injury.

*4.2.2 Scenarios*

(1) *Scenario 1*: There is no reported injury. The manufacturer has been in operation for 4 years and is from a country with a poor safety record for consumer electrical appliances. The manufacturer also has a 'low' customer satisfaction rating, and there are no changes in product design (i.e. product appearance is the same as previous similar products).

(2) *Scenario 2*: There has been 1 reported major injury. The manufacturer has been in operation for 20+ years and is from a country with a very good safety record for consumer electrical appliances. The manufacturer also has a 'very high' customer satisfaction rating, and there are major improvements in product design.

*4.2.3 Method*

Using our BN model, we estimate the probability distribution of hazard per demand for the new uncertified kettle by assigning priors to the following nodes: 1) number of demands tested 2) number of hazards observed during testing 3) testing strategy. These priors are based on testing data for 3 similar kettles. We assume that the data indicate that testing was 'typical of normal use' and the number of demands for the kettles are in the range of 2000 – 2500 with 1 hazard observed. Since we are uncertain about the 'true' number of demands at which the hazard will appear for this particular kettle, we set the lower and upper bounds for the distribution for the number of demands tested as 2000 and 2500 respectively, and we set the number of hazards observed to 1. We then combine the estimated probability distribution of hazard per demand with the manufacturer process information such as country of origin, to estimate the 'true' probability distribution of the hazard per demand.

We calculate the probability distribution of hazard occurrence by combining the user behaviour with the estimated probability distribution of hazard per demand. We do this by assigning priors to the following nodes: 1) particular product usage 2) number of demands 3) years in use. Since we assume that this kettle will be used on average 100 times, we use this value as the mean of the distribution for the number of demands. Also, since we are uncertain about the consumer behaviour during use, we assume that the kettle is used as intended 90% of the times with major and minor deviations of 7% and 3% respectively based on the data for similar kettles. We then calculate the probability distribution of the kettle causing a major and minor injury given the estimated probability distribution of hazard occurrence and the estimated probability distributions of the uncontrolled hazard causing a major and a minor injury respectively. We assume that the probability of the uncontrolled hazard causing a major and a minor injury are 0.1 and 0.2 respectively, and the probability of the control to stop the hazard causing an injury is 0.5.

Finally, we estimate the number of product instances causing major and minor injuries and the overall product risk for the kettle. We assume that there are approximately 50000 – 100000 product instances based on data for similar kettles. Since we are uncertain about the 'true' number of product instances, we set the lower and upper bounds for distribution of the number of product instances as 50000 and 100000 respectively.

Our BN model then predicts the number of product instances causing major and minor injuries and the risk of the new uncertified kettle.

*4.2.4 Results*

(1) *Scenario 1***:** Our BN model shown in Figure 11 learns that the risk level for the new uncertified kettle is 'very high' with some uncertainty. Our BN model also calculates that the mean probability of a major injury is 0.005 and a minor injury is 0.01. It calculates the mean probability of hazard per demand and hazard occurrence as 0.001 and 0.1, respectively. It also estimates that the mean number of major and minor injuries for product instances in the range of 50000-100000 is 375 and 750, respectively. Finally, the BN model shows that the risk tolerability distribution for the kettle is centered at 'low' given a 'medium' utility and recommends a Government intervention such as recall with some uncertainty.

(2) *Scenario 2:* Our BN model shown in Figure 12 learns that the risk level for the new uncertified kettle is 'very low' when the number of product instances causing a major injury is 1. Our BN model also calculates the mean probability of a major and minor injury using Bayes inference as 0.00004 and 0.00009, respectively. It also estimates that the mean number of minor injuries is 6, and the mean probability of the hazard per demand and hazard occurrence to 0.00009 and 0.0009, respectively. Finally, the BN model shows that the risk tolerability distribution for the kettle is in the range of 'high' to 'very high' given a 'medium' utility and recommends no Government intervention with little uncertainty.

Note that, although in Scenario 1 there were no reported injuries from using the product (whereas there was such a report in Scenario 2) the method uses prior and process information to produce recommendations to intervene in Scenario 1 but not Scenario 2.

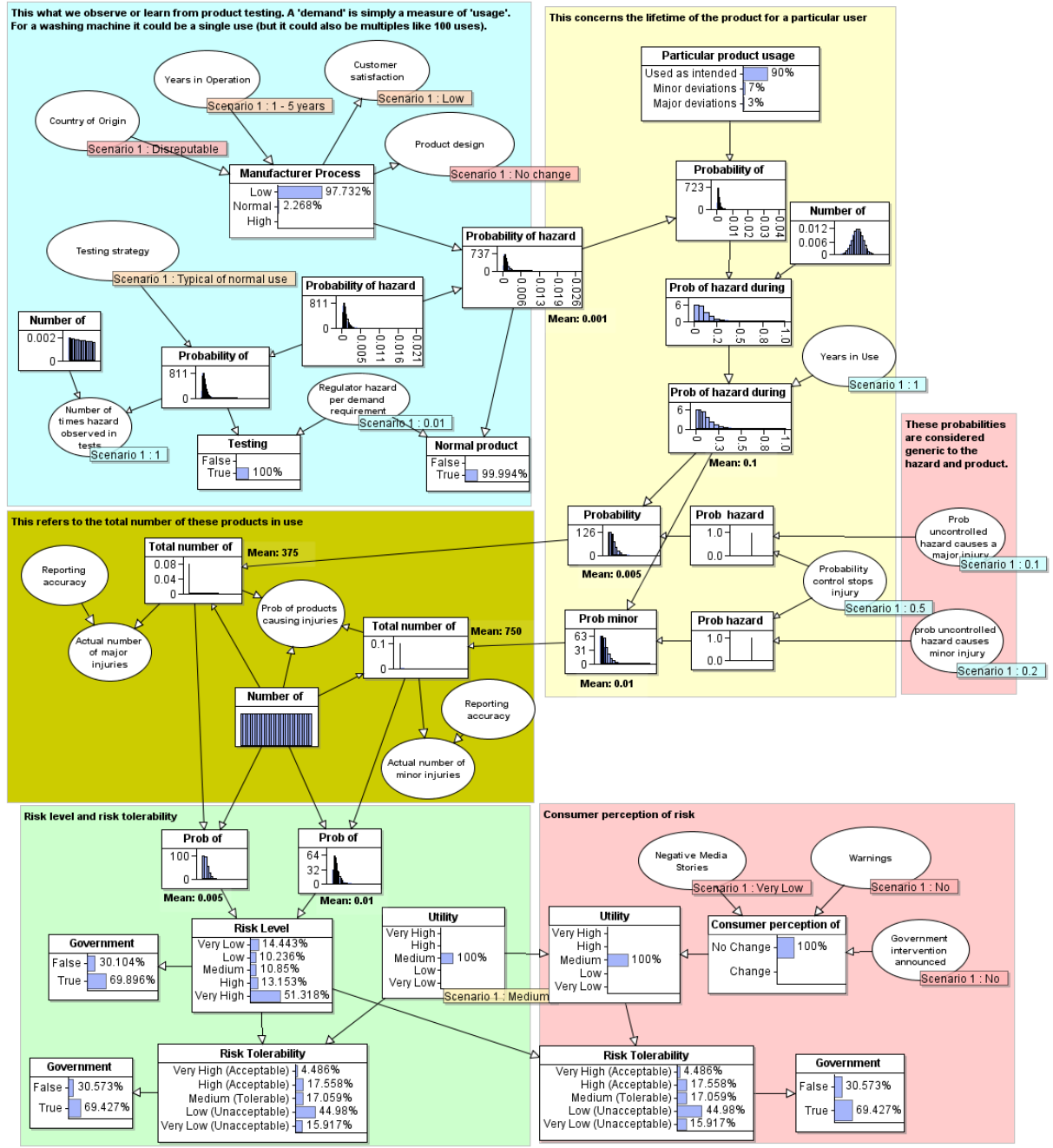

Figure 11. BN model for a new uncertified Kettle Scenario 1

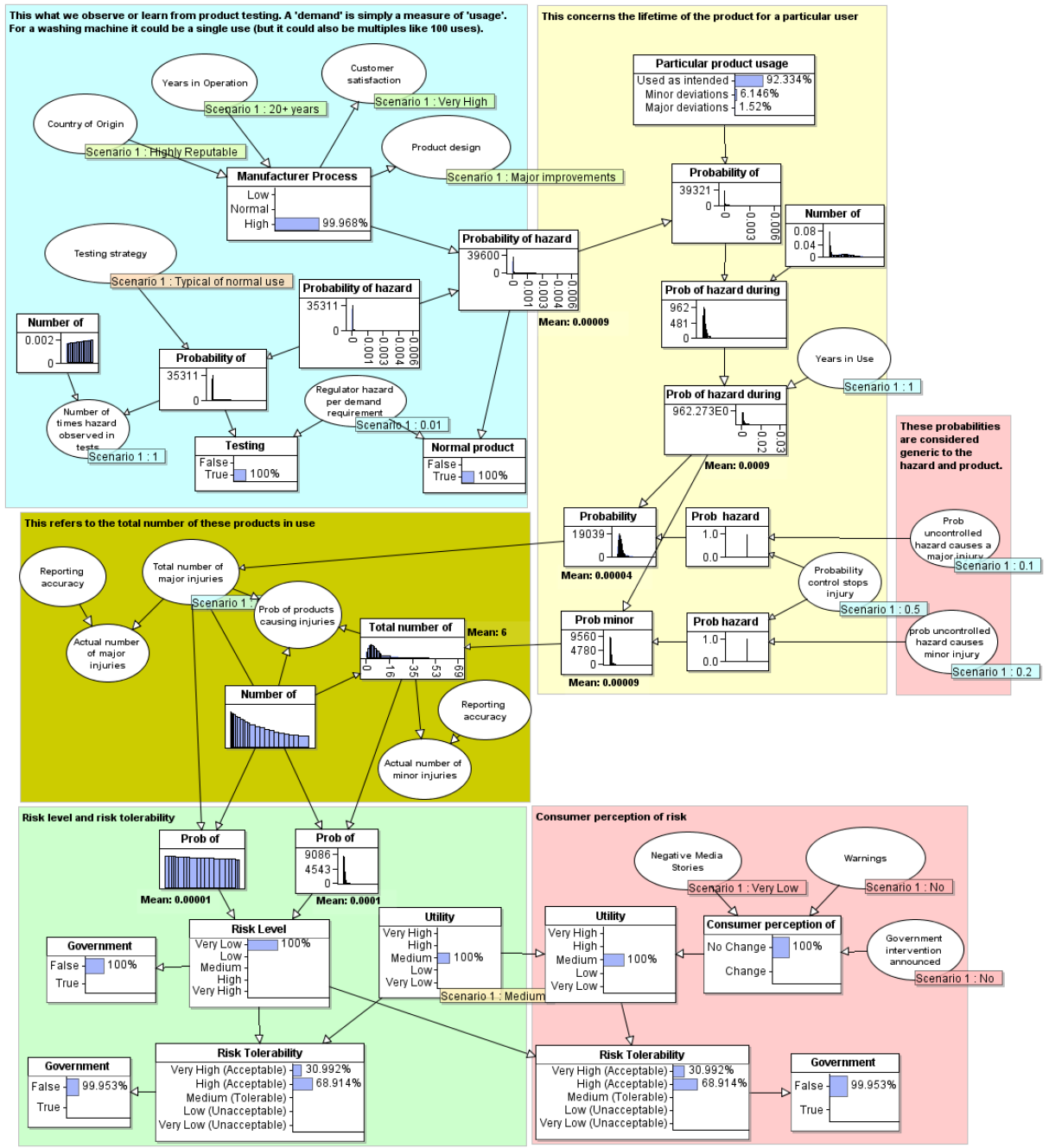

Figure 12. BN model for a new uncertified Kettle Scenario 2

## 5. Conclusion

While we believe the proposed BN approach is a more suitable approach for systematic product risk assessment than approaches like RAPEX, it is important to note that it can also complement RAPEX based assessments. For instance, since the BN approach estimates product risk using additional parameters such as product usage data and manufacturer process information, it can be used in the interim to validate RAPEX risk assessments. In situations where it will neither be feasible nor possible to get any extensive data from testing or injury databases, the BN approach can be used to estimate product risk due to its ability to handle incomplete data, combine objective and subjective evidence and revise risk estimates given new data.

Future work should use behavioural insights to determine consumer behaviour during a particular product lifetime and their perception of risk given a Government intervention, e.g. recall. This will inform our BN model and improve product risk estimates.

## 6. Acknowledgements

This work was supported by the UK Government Department for Business, Energy and Industrial Strategy, Office for Product Safety and Standards (OPSS) and Agena Ltd.

## 7. Disclosure Statement

Norman Fenton and Martin Neil are Directors of Agena Ltd.

RPA. 2006. *Establishing a Comparative Inventory of Approaches and Methods Used by Enforcement Authorities for the Assessment of the Safety of Consumer Products Covered by Directive 2001/95/EC on General Product Safety and Identification of Best Practices*. https://rpaltd.co.uk/uploads/report_files/j497-consumer-products.pdf.

Russo, Jean Nicola, Thomas Sproesser, Frédéric Drouhin, and Michel Basset. 2017. "Risk Level Assessment for Rear-End Collision with Bayesian Network." *IFAC-PapersOnLine*. doi:10.1016/j.ifacol.2017.08.2062.

Spohn, Wolfgang. 2008. "Bayesian Nets Are All There Is to Causal Dependence." *Causation, Coherence, and Concepts*, no. June 2001: 99–111. doi:10.1007/978-1-4020-5474-7_4.

Suh, Jungdae. 2017. "Development of a Product Risk Assessment System Using Injury Information in Korea Consumer Agency." *Journal of Digital Convergence* 15 (4): 181–190. doi:10.14400/jdc.2017.15.4.181.

Wang, Jiali, Martin Neil, and Norman Fenton. 2020. "A Bayesian Network Approach for Cybersecurity Risk Assessment Implementing and Extending the FAIR Model." *Computers and Security*. doi:10.1016/j.cose.2019.101659.

Weber, P., G. Medina-Oliva, C. Simon, and B. Iung. 2012. "Overview on Bayesian Networks Applications for Dependability, Risk Analysis and Maintenance Areas." *Engineering Applications of Artificial Intelligence* 25 (4). Elsevier: 671–682. doi:10.1016/j.engappai.2010.06.002.

Zhang, Limao, Xianguo Wu, Miroslaw J. Skibniewski, Jingbing Zhong, and Yujie Lu. 2014. "Bayesian-Network-Based Safety Risk Analysis in Construction Projects." *Reliability Engineering and System Safety*. doi:10.1016/j.ress.2014.06.006.

**Tables caption list**

- Table 1. Injury severity level and associated medical intervention.

**Figures caption list**

- Figure 1. Bayesian network model for Product risk assessment.
- Figure 2. BN fragment showing no change in consumer perception of risk.
- Figure 3. BN fragment showing the change in consumer perception of risk.
- Figure 4. BN fragment showing the mean number of product instances causing major and minor injuries.
- Figure 5. BN fragment showing the probability distribution of hazard per demand for a particular product with a testing strategy typical of normal use.
- Figure 6. BN fragment showing the probability distribution of hazard per demand for a particular product with a poor testing strategy.
- Figure 7. BN Model for Teddy bear Scenario 1.
- Figure 8. BN Model for Teddy bear Scenario 2.
- Figure 9. RAPEX results for Teddy bear Scenario 1.
- Figure 10. RAPEX results for Teddy bear Scenario 2.
- Figure 11. BN model for a new uncertified Kettle Scenario 1.
- Figure 12. BN model for a new uncertified Kettle Scenario 2.